\documentstyle[12pt]{article}
\newcommand{\Frac}[2]%
{{\textstyle \frac{\mbox{\footnotesize $#1$}\rule[-0.9mm]{0mm}{1mm}}%
{\mbox{\footnotesize $#2$}\rule{0mm}{3.1mm}}}}
\begin{document}
\begin{titlepage}
\vspace*{-12 mm}
\noindent
\begin{flushright}
\begin{tabular}{l@{}}
TK--96--29 \\
hep-ph/9703254 \\
\end{tabular}
\end{flushright}
\vskip 12 mm
\begin{center}
{\large \bf The Drell-Hearn-Gerasimov Sum-Rule in QCD
}\\[14 mm]
{\bf S.D. Bass}\footnote{sbass@pythia.itkp.uni-bonn.de}
\\[10mm]   
{\em Institut f\"{u}r Theoretische Kernphysik, 
Universit\"{a}t Bonn,\\
Nussallee 14--16, D-53115 Bonn, Germany}\\[5mm]
\end{center}
\vskip 10 mm
\begin{abstract}
\noindent
Photoproduction spin sum-rules offer a new window on 
the spin structure of the nucleon that complements 
the information we can learn from polarised deep 
inelastic scattering experiments. 
We review the theory and present status of 
the 
Drell-Hearn-Gerasimov sum-rule in QCD, emphasising the
possible relation between 
the present ``discrepancy'' 
in this sum-rule and the nucleon's strangeness magnetic moment.
In the case of an elementary electron or photon target (say at 
the NLC) the Drell-Hearn-Gerasimov sum-rule provides a test 
for physics 
beyond the minimal Standard Model.

\end{abstract}
\end{titlepage}
\renewcommand{\labelenumi}{(\alph{enumi})}
\renewcommand{\labelenumii}{(\roman{enumii})}
\newpage
\section {Introduction} 

The Drell-Hearn-Gerasimov sum-rule \cite{dhg} for spin dependent 
photoproduction
relates the difference of the two cross-sections for the absorption
of a real photon with spin anti-parallel $\sigma_A$ and parallel 
$\sigma_P$ to the target spin to the square of the anomalous magnetic 
moment of the target, viz.
\begin{equation}
({\rm DHG}) \equiv
- {4 \pi^2 \alpha S \kappa^2 \over m^2} = 
\int_{\nu_{th}}^{\infty} {d \nu \over \nu} (\sigma_A - \sigma_P)(\nu)
\end{equation}
Here $S$ and $m$ are the spin and mass of the target; $\kappa$ is 
the anomalous magnetic moment in units of (${eS \over m}$)  where
$e$ is 
the electric charge.
The sum-rule follows from the very general principles of 
causality, unitarity, Lorentz and electromagnetic gauge invariance
and one assumption:
that we can use an unsubtracted dispersion relation for the spin
dependent part $f_2(\nu)$ of the forward Compton amplitude $f(\nu)$.
Modulo the no-subtraction hypothesis, 
the Drell-Hearn-Gerasimov sum-rule is valid for a target of arbitrary
spin $S$,
whether elementary or composite \cite{brod69}.

In the Standard Model the anomalous magnetic moment
$\kappa = {1 \over 2}(g-2)$   
of an elementary target starts at ${\cal O}(\alpha)$.
It follows that the left hand side of the sum-rule Eq.(1) starts at 
${\cal O}(\alpha^3)$. 
The Drell-Hearn-Gerasimov integral vanishes 
in the classical tree approximation $\sigma \sim {\cal O}(\alpha^2)$
for the
$2 \rightarrow 2$ 
processes
$\gamma a \rightarrow bc$ 
where $a$ is either a
real lepton, quark, photon, gluon 
or 
elementary Higgs target,
viz.
\begin{equation}
\int_{\nu_{th}}^{\infty} {d \nu \over \nu} \Delta \sigma_{\rm Born} 
= 0.
\end{equation}
This result was discovered for a charged lepton target by Altarelli, 
Cabibbo and Maiani
\cite{alt},
and generalised to the other targets
by Brodsky and Schmidt \cite{brodsc}.
Any deviation from the sum-rule Eq.(2) in processes such as
$\gamma e \rightarrow W \nu$ and $\gamma \gamma \rightarrow 
W W$ at
the Next Linear Collider (NLC) would be evidence for physics 
beyond
the minimal Standard Model
such as virtual corrections from SUSY or composite structure 
of the quarks or vector bosons \cite{brodsc, drell, rizzo}.

In QCD the Drell-Hearn-Gerasimov sum-rule provides an important 
constraint on the spin structure of the composite nucleon which
compliments the Bjorken \cite{bj} and Ellis-Jaffe \cite{ej} 
sum-rules of high $Q^2$ 
polarised deep
inelastic scattering.
These high $Q^2$ sum-rules relate the first moment of 
the nucleon's 
first spin structure function $g_1$ to the scale invariant axial 
charges of the target nucleon [7-13].
Polarised deep inelastic scattering experiments 
at CERN and SLAC have verified 
the Bjorken sum-rule for the iso-vector part of $g_1$
to within 8\% \cite{exptbj,abfr}.
They have also revealed a four standard deviations violation 
of OZI in the flavour singlet axial charge $g_A^0|_{\rm inv}$ 
\cite{expt, badelek}.
This discovery has inspired many would-be theoretical 
explanations involving the axial anomaly [18-23] and chiral
soliton models [24] -- for recent reviews see \cite{altr,cheng}.

A longstanding puzzle in (pre-QCD) spin physics is the ``discrepancy''
between the {\it inclusive} Drell-Hearn-Gerasimov sum-rule for a 
nucleon target and the ``measured'' {\it exclusive} one and two pion
photoproduction contributions to the Drell-Hearn-Gerasimov sum-rule.
I. Karliner \cite{ikar}, following earlier work by Fox and Freedman
\cite{ffox}, carried out a
multipole analysis of (unpolarised) single pion photoproduction data
to obtain the contribution of this exclusive channel,
which seems to be well saturated below $\nu_{\rm LAB} \sim 1.2$GeV,
to the inclusive 
sum-rule.
When she combined this analysis with an estimate of the two pion 
background, 
Karliner found very close agreement between these exclusive 
``measurements'' and the isoscalar part of the fully inclusive 
sum-rule, 
but not 
the isovector
part, which differed both in sign and by a factor of 2.
The theoretical predictions for the isoscalar and isovector parts
of the DHG integral 
$({\rm DHG})_{(I=0,1)}^{\rm inclusive}$
and
Karliner's results $({\rm DHG})_{(I=0,1)}^{\rm \pi}$
are:
\begin{eqnarray}
({\rm DHG})_{I=0}^{\rm inclusive} = -219{\rm \mu b},   \ \ \ \ \ \ 
({\rm DHG})_{I=0}^{\rm \pi} =       -222{\rm \mu b}   \\ \nonumber
\\ \nonumber
({\rm DHG})_{I=1}^{\rm inclusive} = +15{\rm \mu b},   \ \ \ \ \ \ 
({\rm DHG})_{I=1}^{\rm \pi} =       -39{\rm \mu b}.   \\ \nonumber
\end{eqnarray}
No error was quoted for the ``experimentally determined'' 
integrals $({\rm DHG})_{(I=0,1)}^{\rm \pi}$, which were evaluated 
up to an ultraviolet cutoff $\nu_{\rm LAB}=1.2$GeV. 
Whilst one should not worry that an exclusive channel does not
saturate an inclusive sum-rule,  this result is very curious 
in that we normally expect the isovector contribution to an
inclusive sum-rule to saturate faster with increasing energy
$\nu$ than the isoscalar contribution. 
(Pomeron-like exchanges which govern the high-energy behaviour 
of $f(\nu)$ are isoscalar.)

The first direct test of the Drell-Hearn-Gerasimov sum-rule will 
be made in experiments which are planned or underway at the 
ELSA, GRAAL, LEGS and MAMI facilities.
Whilst we wait for the results of these experiments it is
worthwhile to review what is known and what is not known 
about the spin structure of the nucleon at photoproduction.
This Brief Review is about the Drell-Hearn-Gerasimov sum-rule 
in QCD. 
We refer to Drechsel \cite{drec} for an excellent review of 
nuclear physics aspects.

The structure of this article is as follows. 
In Section 2 we review the derivation of photoproduction 
sum-rules with emphasis on the validity of the no-subtraction 
hypothesis. In Section 3 we discuss the consequences of
the Drell-Hearn-Gerasimov sum-rule for our understanding
of nucleon structure.
We then give an overview of the present status of 
the Karliner ``discrepancy'' (Section 4) and possible resolutions 
of this ``discrepancy'' based on strange quark dynamics (Section 5)
and vector meson dominance arguments (Section 6).
Section 6 also reviews the $Q^2$ dependence of 
$(\sigma_A - \sigma_P)$ from photoproduction to polarised 
deep inelastic scattering.
Finally, we summarise the physics that can be learnt in future
polarised photoproduction experiments.

\section {The Drell-Hearn-Gerasimov Sum-Rule}

To understand the derivation of the Drell-Hearn-Gerasimov
Sum-Rule we start with the forward Compton scattering
amplitude
\begin{equation}
f (\nu) =
\chi_f^* \Biggl[
f_1 (\nu) {\vec \epsilon}_f^*.{\vec \epsilon}_i + i f_2 (\nu)
{\vec \sigma}.({\vec \epsilon}_f^* {\rm x} {\vec \epsilon}_i )
\biggr] \chi_i.
\end{equation}
{\it If} 
$|f_2 (\nu)| \rightarrow 0$ for complex $\nu \rightarrow \infty$,
{\it then}
causality allows us to write an unsubtracted dispersion relation
for the spin dependent part of $f(\nu)$.
Using the Optical Theorem (unitarity) to rewrite ${\rm Im}f_2$ in 
terms of the photoproduction cross-sections and crossing symmetry
\begin{eqnarray}
f_2(\nu) = - f_2^*(-\nu)
\end{eqnarray}
we obtain
\begin{equation}
{\rm Re} f_2 (\nu) = {\nu \over 4 \pi^2}
\int_{\nu_{th}}^{\infty} d \nu' {\nu' \over \nu'^2 - \nu^2}
( \sigma_A - \sigma_P )(\nu').
\end{equation}
Photoproduction spin sum-rules are obtained by taking the odd 
derivatives (with
respect to $\nu$) of Re$f_2(\nu)$ in Eq.(6),
viz.
$\biggl(
{\partial \over \partial \nu} \biggr)^{2n+1}$Re$f_2(\nu)$, 
and then evaluating
the resulting expression at $\nu =0$.
For small photon energy $\nu \rightarrow 0$, we use the
low energy theorems \cite{low}:
\begin{equation}
f_1 (\nu) \rightarrow - {\alpha \over m} + 
({\overline \alpha}_N + {\overline \beta}_N) \nu^2
+ {\rm O}(\nu^4)
\end{equation}
and
\begin{equation}
f_2 (\nu) \rightarrow
- {\alpha \kappa^2_N \over 2 m^2} \nu + \gamma_N \nu^3 
+ {\rm O}(\nu^5)
\end{equation}
which follow from Lorentz and electromagnetic gauge 
invariance.
Here 
$\gamma_N$,  
${\overline \alpha}_N$ and ${\overline \beta}_N$ 
are the
spin, and electric and magnetic
Rayleigh polarisabilities of the nucleon respectively.

The Drell-Hearn-Gerasimov sum-rule is derived when we 
evaluate the first derivative
($n=0$) of Re$f_2(\nu)$ in Eq.(6) at $\nu =0$, viz.
\begin{equation}
\int_0^{\infty} {d \nu' \over \nu'}
\Biggl ( \sigma_A - \sigma_P \Biggr )(\nu')
= - { 2 \pi^2 \alpha \over m^2 } \kappa^2_N.
\end{equation}
The third derivative of Re$f_2 (\nu)$ yields a second 
sum-rule for the spin
polarisibility $\gamma_N$
\begin{equation}
\int_0^{\infty} {d \nu' \over \nu'^3}
\Biggl (\sigma_A - \sigma_P \Biggr)(\nu')
=
{1 \over 4 \pi^2} \gamma_N.
\end{equation}
These photoproduction sum-rules have a parallel in high $Q^2$ 
deep inelastic scattering where the odd moments
of the spin dependent structure function $g_1$
project out the nucleon matrix elements of gauge-invariant 
local operators, each of
which is multiplied
by a radiative Wilson coefficient.

The no-subtraction hypothesis is the only ``QCD input" 
to the derivation of the sum-rule. 
How good is it ?

Discounting for the moment any fixed pole contribution,
${\rm Im}f_2$ and $f_2$ have the same 
$\nu \rightarrow \infty$
behaviour
in the Regge analysis of the high-energy behaviour of 
spin dependent hadronic scattering processes \cite{coll}.
Regge arguments tell us that the Drell-Hearn-Gerasimov
integral converges.

It is well known \cite{peie,heim} that the isotriplet
contribution to
$(\sigma_A - \sigma_P)$
behaves as
\begin{equation}
\Biggl( \sigma_A - \sigma_P \Biggr)_3 (\nu) 
        \sim \nu^{\alpha_{a_1} - 1}, 
\ \ \ \ \nu \rightarrow \infty
\end{equation}
where $\alpha_{a_1}$ is the intercept of the $a_1$ 
Regge trajectory 
($-0.5 \leq \alpha_{a_1} \leq 0$ \cite{ek}) .
The isosinglet contribution is still somewhat controversial 
and depends 
on
the Lorentz structure
of the short range part of the exchange potential \cite{clos1}.
The soft pomeron dominates the large energy ($\sqrt{s} > 10$GeV) 
behaviour of the spin averaged cross section 
$(\sigma_A + \sigma_P)$ \cite{pvl1}
but does not contribute to 
$(\sigma_A - \sigma_P)$ \cite{heim}.
However, the physics which generates the soft pomeron 
can 
contribute to $(\sigma_A - \sigma_P)$.
If the pomeron were a scalar, then we would find
\begin{equation}
\Biggl(\sigma_A - \sigma_P \Biggr)_0^{[1]} 
       \sim 0, \ \ \ \ \nu \rightarrow \infty.
\end{equation}
In the Landshoff-Nachtmann approach \cite{pvl1,pvl2}, 
the soft pomeron is modelled by the exchange of 
two non-perturbative gluons
and transforms as a $C=+1$ vector potential 
with a correlation length of about 0.1fm.
This two non-perturbative gluon exchange gives a 
contribution \cite{sbpl, clos1}
\begin{equation}
\Biggl( \sigma_A - \sigma_P \Biggr)_0^{[2]} (\nu) 
        \sim { \ln \nu \over \nu},
\ \ \ \ \nu \rightarrow \infty.
\end{equation}
It has also been suggested \cite{kuti}
that there may be a negative-signature, two-pomeron
cut contribution to $(\sigma_A - \sigma_P)$
\begin{equation}
\Biggl(\sigma_A - \sigma_P \Biggr)_0^{[3]} (\nu) 
       \sim { 1 \over \ln^2 \nu },
\ \ \ \ \nu \rightarrow \infty.
\end{equation}
It is an important challenge for future high-energy 
photoproduction experiments to determine the 
strength of the three possible large $\nu$ contributions 
in Eqs.(12-14).
One could then make contact with the small $x$ behaviour 
of the nucleon's first spin structure function $g_1$ 
which is measured in polarised deep inelastic scattering.
The soft Regge contributions form a baseline for 
investigations of DGLAP \cite{dglap} and BFKL \cite{bfkl} 
small $x$ behaviour in $g_1$.
Each of the possible contributions in Eqs.(11-14) give a 
convergent integral in Eqs.(6,9).
If the short range exchange potential transformed as 
an axial-vector, then the Drell-Hearn-Gerasimov integral 
would not have converged: the sea would be generated all 
with the same polarisation and
$(\sigma_A - \sigma_P)$
would grow to reach the Froissart bound for the spin 
averaged cross section
\begin{equation}
\Biggl(\sigma_A + \sigma_P \Biggr) \sim \ln^2 \nu, \ \ \ \ 
\nu \rightarrow \infty.
\end{equation}
However, this is not QCD.
(In QCD the $qqg$ vertex is a vector coupling -- see eg. \cite{Muta}.)

Given that the Drell-Hearn-Gerasimov integral converges, 
the only way that the sum-rule can fail is
if there exists a $J=1$ Regge fixed pole so that 
${\rm Im}f_2 \rightarrow 0$ does vanish 
and
${\rm Re}f_2$ does not vanish at asymptotic photon energy 
$\nu \rightarrow \infty$.
In this case the condition
$|f_2 (\nu)| \rightarrow 0$ as $\nu \rightarrow \infty$ would
not be fulfilled
and we would find a constant spin flip 
(subtraction constant) correction to the sum-rule.
There is both theoretical and phenomenological evidence against 
such
a fixed pole correction to (DHG), Eq.(1).

Fixed pole contributions \cite{coll, dash} to current-hadron
scattering amplitudes like $f(\nu)$ are, in principle, 
allowed by current algebra.
They are not allowed in hadron-hadron amplitudes.
Consider a photon scattering from a (composite) target, 
the structure 
of which is described by a field theory with coupling constant 
$g$.
Where they exist, arguments in favour of fixed poles in $f_1$
yield 
contributions which start at ${\cal O}(1)$ in the transverse
cross-section \cite{brod72}
and ${\cal O}(g^2)$ in the longitudinal cross-section \cite{bgj}.
Explicit calculation \cite{alt,brodsc,pfeil} shows that both 
sides of the Drell-Hearn-Gerasimov sum-rule vanish at 
${\cal O}(\alpha_s \sim g_s^2)$ in perturbative QCD, 
in agreement with the sum-rule Eq.(2).
There is no constant spin flip correction to the sum-rule at
either ${\cal O}(1)$ or ${\cal O}(\alpha_s)$ in perturbative
QCD.
This result suggests that any fixed pole contribution is most
likely a non-perturbative effect.

It was suggested in \cite{gold} that an isovector $J=1$ fixed 
pole might 
resolve the ``discrepancy'' 
between 
$({\rm DHG})_{I=1}^{\rm inclusive}$ and
$({\rm DHG})_{I=1}^{\pi}$.
No suggestion was offered why the isoscalar partner 
of such a  would-be fixed pole should be suppressed.
As emphasised by Heimann \cite{heim}, an isovector $J=1$ 
fixed pole would 
also induce a violation of Bjorken's sum-rule in polarised
deep inelastic scattering.
The Bjorken sum-rule is presently verified to within 8\% 
\cite{exptbj, abfr}.

Henceforth, we accept the Drell-Hearn-Gerasimov sum-rule as valid.

\section {Properties of the DHG sum-rule}

The forward Compton amplitude $f(\nu)$ in photon-nucleon scattering
has charge parity $C=+ 1$.
The cross-sections for spin dependent photoproduction (on the right
hand side of the Drell-Hearn-Gerasimov sum-rule) receive 
contributions from OZI violating processes where the photon couples 
to the target nucleon via a $C=+1$, colour neutral, gluonic intermediate 
state in the $t$ channel, 
for example two gluon exchange.

The anomalous magnetic moment $\kappa_N$ which appears on the left 
hand side of the Drell-Hearn-Gerasimov sum-rule has charge parity 
$C= -1$. It is measured in the nucleon matrix element of the conserved 
vector current operator which is not renormalised. 
(It is renormalisation group invariant and does not mix with 
any gluonic operators under renormalisation.)
Furry's theorem tells us that $\kappa_N$ receives no contribution 
from processes where the photon couples to a quark loop which, in 
turn, couples to an even number of gluons.
The axial anomaly \cite{adler, bell} does not contribute to $\kappa_N$.
OZI violation can arise in $\kappa_N$ via the photon's coupling
to the $\phi$(1020), which, in turn, can couple to the 
three quark ``valence nucleon'' via a $C=-1$, colour neutral, 
three gluon intermediate state.

The charge parity properties of $\kappa_N$ and $f_2$ on the left 
and right hand sides of the sum-rule Eq.(1) imply that processes 
which contribute to $C= +1$ observables but not to matrix 
elements of
the conserved $C=-1$ vector current
cancel 
in the logarithmic 
Drell-Hearn-Gerasimov integral for the difference of 
the two spin dependent photoabsorption cross-sections \cite{moi}.
In the absence of a $J=1$ fixed pole, the very general 
principles of causality, unitarity, Lorentz and electromagnetic 
gauge invariance
imply that
the contribution to the Drell-Hearn-Gerasimov integral 
from 
OZI violating exchanges which do not contribute to $\kappa_N$
is either zero or infinite
(so that the integral would not converge and one would 
 need a subtraction).
In the infinite scenario the sea and glue would 
be generated all 
with one polarisation, which is not QCD.

In contrast to the Drell-Hearn-Gerasimov sum-rule, the Spin
Polarisability
sum-rule involves only $C=+ 1$ observables and is sensitive
to all $C=+1$ OZI violating exchanges in the $t$ channel.

It is interesting to compare the physics of Drell-Hearn-Gerasimov 
with the 
spin averaged cross section.
The negative sign of the Thomson term $- {\alpha \over m}$ in the
low energy theorem for $f_1(\nu)$, Eq.(7), 
means that we cannot write an unsubtracted 
dispersion relation
for the spin averaged cross-section \cite{gilm}.
(Cross-sections are non-negative!)
Furthermore, the Thomson term involves the square of 
the forward matrix element of the $C=-1$ vector current 
(the Dirac 
 form-factor evaluated at zero momentum transfer),
which measures the number of valence (current or constituent) 
quarks in the nucleon.
Even without the sign, $C=+1$
OZI violating processes involving two gluon exchange 
(like Pomeron exchange) will contribute to the total 
spin-averaged cross-section
so that any integral sum-rule
for this cross-section
must involve at least one $C= +1$ observable.

The Drell-Hearn-Gerasimov sum-rule also tells us that
the contribution to the anomalous magnetic moment of 
a massive fermion due to internal structure
vanishes in the limit 
$m_f r \rightarrow 0$, where $m_f$ is the mass and $r$ is 
the radius of the fermion \cite{drell}.
Let $m^{*} \gg m_f$ be a scale that
represents the
internal bound-state structure of the fermion 
($m^{*} \rightarrow \infty$ in the limit that
$r \rightarrow 0$).
In addition to pure QED contributions, the 
Drell-Hearn-Gerasimov integral will 
receive new resonance or continuum 
contributions beyond the threshold
$s_{th} = m^{*2}$
associated with the internal structure of the fermion.
Such contributions yield
\begin{equation}
\biggl(\sigma_A - \sigma_P \biggr) 
\sim 
\alpha \ {\pi \over m^{*2}} f({m^{*2} \over s^2})
\end{equation}
and a contribution to the anomalous magnetic
moment
\begin{equation}
\delta \kappa^{\rm (non-QED)} \sim {m_f \over m^{*}}
\end{equation}
which vanishes in the zero radius limit.
If we use the Drell-Hearn-Gerasimov sum-rule as a 
probe for physics beyond the minimal Standard 
Model in an NLC experiment,
then $m^*$ represents the scale of ``new physics''.
In QCD
Eq.(17) provides an important constraint on quark model
expressions for the anomalous magnetic moment.
In the MIT and Cloudy Bag Models the nucleon's 
anomalous magnetic moment
is proportional to the confinement radius $R$
\cite{cloud}
(see also \cite{schl}).
On the other hand, the non-relativistic quark 
model expression
\begin{equation}
\kappa_N = \sum_q e_q {\sigma_q} {m \over m_q} -1,
\end{equation}
is inconsistent with the Drell-Hearn-Gerasimov
sum-rule
beacuse it contains no information about the radius
of the nucleon.
(In Eq.(18) $e_q$ is the quark charge, $m_q$ is the 
 constituent quark mass and $m$ is the nucleon mass. 
 The $C=-1$ quantity $\sigma_q$ denotes the fraction 
 of the nucleon's spin that is carried by its quarks 
 minus the fraction that is carried by its antiquarks.)

\section {The Experimental Situation }

To understand Karliner's observation of a ``discrepancy'' 
between the inclusive Drell-Hearn-Gerasimov sum-rule and 
the contribution to the sum-rule from exclusive one and 
two pion photoproduction, it is helpful to write the anomalous 
magnetic moment $\kappa_N$ as the sum of its isovector 
$\kappa_V$ and isoscalar $\kappa_S$ parts, 
viz.
\begin{equation}
\kappa_N = \kappa_S + \tau_3 \kappa_V.
\end{equation}
We can then write separate isospin sum-rules:
\begin{eqnarray}
  ({\rm DHG})_{I=0} = ({\rm DHG})_{VV} + ({\rm DHG})_{SS} 
                    &=& - {2 \pi^2 \alpha \over m^2} 
                         (\kappa_V^2 + \kappa_S^2)     \\ \nonumber
  ({\rm DHG})_{I=1} = ({\rm DHG})_{VS}  
                    &=& - {2 \pi^2 \alpha \over m^2} 
                         2 \kappa_V \kappa_S.
\end{eqnarray}
The physical values of the proton and nucleon anomalous 
magnetic moments $\kappa_p = 1.79$ and $\kappa_n = -1.91$
correspond to
\begin{eqnarray}
\kappa_S = -0.06 \\ \nonumber
\kappa_V = +1.85
\end{eqnarray}
Since $\Bigl( {\kappa_S \over \kappa_V} \Bigr) \simeq - {1 \over 30}$, 
it follows that $({\rm DHG})_{SS}$ is negligible compared 
to $({\rm DHG})_{VV}$.
Indeed, it is safe to regard $({\rm DHG})_{SS}$ 
as negligible compared to any reasonable estimate of the present 
``experimental'' error
on $({\rm DHG})_{I=0}$.
It follows that, first, the isoscalar sum-rule $({\rm DHG})_{I=0}$ 
measures the isovector anomalous magnetic moment $\kappa_V$.
Given this isoscalar measurement,
the isovector sum-rule $({\rm DHG})_{I=1}$ then measures the 
isoscalar anomalous
magnetic moment $\kappa_S$.
If we take the Drell-Hearn-Gerasimov sum-rule as an exact equation 
and just the Karliner data, 
then we can invert this equation to obtain
\begin{eqnarray}
  \kappa_S^{(\pi)} &=& +0.16       \\ \nonumber
  \kappa_V^{(\pi)} &=& +1.86.
\end{eqnarray}
The ``discrepancy'' in the {\it isovector} sum-rule 
translates into a ``discrepancy'' in the {\it isoscalar} 
anomalous magnetic moment $\kappa_S$.
(Note that 
$ \biggl( {\kappa_S \over \kappa_V} \biggr)^{(\pi)} \simeq 
  - {1 \over 12} $
so that 
$({\rm DHG})_{SS}^{(\pi)} \ll ({\rm DHG})_{VV}^{(\pi)} $ 
with these values of $\kappa$.)

Karliner's evaluation of $({\rm DHG})_{(I=0,1)}^{(\pi)}$,
and hence $\kappa_{(V,S)}^{(\pi)}$, emerged as a
delicate 
cancellation of individual multipole contributions, 
which we list in Table 1.
(We refer to \cite{drec, drec92} for the definitions of 
 these resonances and multipoles.)
\begin{table}
\begin{center}
\caption{Multipole Structure of the DHG integral }
\begin{tabular} {cccc}
\\
\hline\hline
\\
Resonance & Multipole & Contribution to 
${\rm (DHG)}_{I=0}$ &
Contribution to ${\rm (DHG)}_{I=1}$  \\
\\
\hline
\\
$P_{33}, {3 \over 2}^{+}$ & $M_{1+} (E_{1+})$ & -271$\mu$b & +23$\mu$b \\
                    & ($\Delta(1232)$ only    & -240$\mu$b & )     \\
$S_{11}, {1 \over 2}^{-}$ & $E_{0+}         $ & +168$\mu$b & -23$\mu$b \\
$P_{11}, {1 \over 2}^{+}$ & $ M_{1-}        $ & + 10$\mu$b & - 3$\mu$b \\
$D_{13}, {3 \over 2}^{-}$ & $E_{2-},M_{2-}  $ & - 67$\mu$b & -13$\mu$b \\
$F_{15}, {5 \over 2}^{+}$ & $E_{3-},M_{3-}  $ & - 10$\mu$b & - 8$\mu$b \\
  &      $2 \pi$ background                   & - 52$\mu$b & -16$\mu$b  \\
\\
\hline
  &      ``experiment''   &                     -222$\mu$b & -39$\mu$b \\
  &      DHG              &                     -219$\mu$b & +15$\mu$b \\
\hline\hline
\end{tabular}
\end{center}
\end{table}
A more recent multipole analysis has been performed by Workman 
and Arndt \cite{workm} using an updated data set for single 
pion photoproduction up to $\nu_{\rm LAB} = 1.7$GeV.
They obtained
$({\rm DHG})_{I=1}^{({\rm \pi, WA})} = -225\mu$b and
$({\rm DHG})_{I=0}^{({\rm \pi, WA})} = - 65\mu$b 
\cite{workm} (see also \cite{spinp})
by combining
their new multipole analysis with Karliner's estimate of 
the two pion background (up to $\nu_{\rm LAB} = 1.2$GeV).
The main difference between the two analyses is in the
single pion photoproduction data
between 400MeV$< \nu_{\rm LAB} <$600MeV. 
If taken at face value, this new evaluation
of the Drell-Hearn-Gerasimov integral would appear to increase
the ``discrepancy'' in the sum-rule.
However, it is important to bear in mind that different 
ultraviolet cut-offs were used to evaluate the one and two pion 
contributions
to $({\rm DHG})_{(I=0,1)}^{({\rm \pi, WA})}$.
Furthermore, and most importantly, both the Karliner \cite{ikar}
and Workman and Arndt \cite{workm} determinations 
of $({\rm DHG})$ 
are indirect ``measurements'' extracted from unpolarised data.
We eagerly await the results of the ELSA, GRAAL, LEGS and MAMI
experiments, 
which will provide the first direct measurements of 
$\sigma_A$ and $\sigma_P$ up to $\nu_{\rm LAB} \simeq 3$GeV.

To resolve Karliner's ``discrepancy'' we are looking 
for non-resonant contributions below $\nu_{\rm LAB} = 1.2$GeV
as well as physics which contributes to $(\sigma_A - \sigma_P)$ 
at 
$\nu_{\rm LAB} > 1.2$GeV 
which can shift $\kappa_S$ by an amount $\simeq -0.22$ whilst 
keeping $\kappa_V$ essentially constant.
There is no elastic contribution to the DHG integral.
In the Introduction we outlined how the DHG sum-rule
can be used to look for physics beyond the minimal Standard
Model in an NLC experiment.
This argument \cite{brodsc} can be extended to a nucleon target
as follows.
The size of an {\it exclusive} channel contribution to the {\it 
inclusive} Drell-Hearn-Gerasimov sum-rule 
measures the contribution of the physics involved in this channel 
to the anomalous magnetic moment with all other physics switched
off.

Now assume that pion photoproduction can be described entirely
by light quark dynamics.
In this scenario, both $\kappa_S^{(\pi)}$ and $\kappa_V^{(\pi)}$ 
do not receive any contribution from the physics which generates 
the nucleon's
strangeness magnetic moment 
\footnote{
For completeness, we note two places where dynamics involving
the strange quark 
can contribute to pion photoproduction.
First, in a full treatment, one should include virtual kaon 
corrections to the $\gamma N \rightarrow N^{*}$ vertices.
These corrections,
which are in general much smaller than the corresponding 
virtual pion corrections, 
are absorbed into the electromagnetic form-factors for these transitions.
Second, 
the prominant role of the $S_{11}(1535)$ 
in eta photoproduction suggests that the wavefunction 
of this resonance contains a significant $\Sigma K$ 
bound state component \cite{weise}. 
We assume here that the pure light-quark component 
of the $S_{11}(1535)$ dominates its contribution 
to pion photoproduction.}.
The physics of the strangeness magnetic moment (Section 5) 
is one possible candidate to fill the Karliner ``discrepancy''.

It is interesting to compare the present status of 
the Drell-Hearn-Gerasimov and Spin Polarisability sum-rules 
\cite{spinp}.
In Table 2 we show the values of the proton and neutron spin 
polarisabilities that
are extracted from the pion photoproduction data in comparison
with the predictions \cite{chpt} of relativistic, one loop chiral 
perturbation theory, $\chi$PT,
with and without the $\Delta$ resonance included.
\begin{table}
\begin{center}
\caption{Present status of the Spin Polarisability sum-rule}
\begin{tabular} {cccc}
\\
\hline\hline
\\
$\gamma_N$ 
  &  Multipoles  &  $\chi$PT (1 loop without $\Delta$)  &
                    $\chi$PT (1 loop with $\Delta$   )  \\
\hline
\\
$\gamma_p$  &  -1.34$\mu$b  &  +2.16$\mu$b  &  -1.50$\mu$b  \\
$\gamma_n$  &  -0.38$\mu$b  &  +3.20$\mu$b  &  -0.46$\mu$b  \\
\hline\hline
\end{tabular}
\end{center}
\end{table}
There is good agreement between the ``data'' and this $\chi$PT 
calculation with the $\Delta$ included.
This result is quite natural.
First, the Spin Polarisability sum-rule is expected to saturate
faster with increasing energy than the 
Drell-Hearn-Gerasimov sum-rule because of the ${1 \over \nu^3}$
denominator in Eq.(10).
The $\Delta$ as the lowest lying nucleon resonance is therefore
expected to play a more prominant role in the sum-rule for
$\gamma_N$ than the sum-rule for the anomalous magnetic moment.
Second, both the ``experimental'' numbers and theoretical 
predictions for $\gamma_N$ in Table 2 
do not include dynamics associated with the strange quark.

\section {Relation to the strangeness magnetic moment }

The nucleon's anomalous magnetic moment is measured in the 
$q^2 \rightarrow 0$
limit of the nucleon's
Pauli form-factor, viz. $\kappa_N = F_2(0)$ 
where
\begin{equation}
<N(p') | j_{\mu}^{\rm em}(0) | N(p)> =
{\overline u}(p') \biggl( \gamma_{\mu} F_1(Q^2) + 
{i \sigma_{\mu \nu} q^{\nu} \over 2 m} F_2(Q^2) \biggr)
u(p)
\end{equation}
and $q = (p-p')$.
We can express $\kappa_p$ as the sum over up, down and 
strange quark
contributions
\begin{equation}
\kappa_p = {2 \over 3} F_2^u(0) - {1 \over 3} F_2^d(0)
         - {1 \over 3} F_2^s(0).
\end{equation}
Motivated in part by the role that intrinsic strangeness 
plays in some theoretical explanations of the EMC spin effect 
and the value
of the pion-nucleon sigma term,
there is presently much 
interest in measuring the size
of $\kappa_s$ with experiments planned at Bates, MAMI and TJNAF.

Theoretical predictions [58-69] of $F_2^s(0)$ are very model 
dependent with results that differ both in sign and order of 
magnitude.
There are two main types of approach.
The first has its starting point H\"ohler et al.'s 
minimal three pole approximation \cite{hohl1} to the spectral 
function for
the iso-scalar Pauli form-factor
\begin{equation}
F_2^{(I=0)}(q^2) = \sum_{i=1}^3 {A_i m_i^2 \over m_i^2 - q^2 }.
\end{equation}
Using vector meson dominance arguments, the first 
pole is taken to be the $\omega(780)$.
H\"ohler et al. \cite{hohl1} found the second pole at
$m_2^2 = 0.97$GeV$^2$  which they identified with the 
$\phi(1020)$.
The $\phi(1020)$ couples to the nucleon with a 
large OZI violation \cite{hohl2} and, in this approach, leads
to a large negative strangeness magnetic moment $F_2^s(0) \simeq -0.26$
\cite{jaffe}.
This result has to be taken with care.
First, Meissner et al. \cite{meis97} have recently emphasised 
the role of a correlated $\pi \rho$ exchange
which is modelled by a single effective $\omega'$ 
pole with mass $m_{\omega'} = 1.1$GeV in the Bonn 
potential for the nucleon-nucleon interaction \cite{holi}.
This effective $\omega'$ is an alternative candidate for 
the second pole $m_2$ in Eq.(25). 
It has the potential to remove the OZI violation found
by saturating
the $m_2$ pole with the $\phi(1020)$.
Secondly, in a full treatment, it is necessary to extend 
the fit to $F_2^{(I=0)}(q^2)$ beyond 
the three pole approximation to include continuum contributions
and to match onto perturbative QCD asymptotics at large $q^2$,
viz.
$F_2 \sim \biggl( {1 \over q^2} \biggr)^3, q^2 \rightarrow \infty$
\cite{brod80}.

The second approach to $F_2^s(0)$ involves calculating
$N \rightarrow K \Lambda$ and $N \rightarrow K \Sigma$ 
loops within a given model.
The first authors to do this were Henley and collaborators
\cite{henl} 
who found $F_2^s(0) \simeq -0.03$ in the Cloudy Bag Model 
\cite{cloud}
with SU(3) couplings.
If SU(3) couplings are used (as in all the present calculations),
then
$\Bigl( {g^2_{K \Sigma N} \over g^2_{K \Lambda N}} \Bigr)_{\rm SU(3)}
 \simeq {1 \over 25}$
and the $N \rightarrow K \Sigma$ loop contributions to $F_2^s(0)$ 
are neglected in a first approximation.
($K \Sigma$ loops will play a more important role if 
phenomenological, SU(3) violating, couplings 
$\Bigl( {g^2_{K \Sigma N} \over g^2_{K \Lambda N}} \Bigr)_{\rm ph}
 \simeq {1 \over 4}$
\cite{sign} are used.)
Musolf and Burkardt \cite{muso} subsequently argued that seagull 
terms can
induce a large negative value of $F_2^s(0) \simeq -0.35$.
Melnitchouk and Malheiro \cite{wall} have recently estimated 
$F_2^s(0)$ using light-cone form-factors which are consistent 
with 
mesonic effects in deep inelastic scattering \cite{shimoda}.
Without seagulls,
they find that these kaon loops contribute between -0.04 and -0.1 
to $F_2^s(0)$.
In a recent quark model calculation, Geiger and Isgur \cite{isgur} 
have included 
the processes
($p \rightarrow (\Lambda K)_{P({1 \over 2})}$, 
$\gamma (\Lambda K)_{P({1 \over 2})} 
\rightarrow
(\Lambda K^*)_{P({1 \over 2})} \rightarrow p$) 
and
($p \rightarrow [\Lambda^*(1405) K]_{S({1 \over 2})}$, 
$\gamma [\Lambda^*(1405) K]_{S({1 \over 2})} 
\rightarrow
[\Lambda^*(1405) K^*]_{S({1 \over 2})} \rightarrow p$)
inside the kaon loop.
They obtain $F_2^s(0) \simeq +0.04$ in their full
calculation and $F_2^s(0) \simeq -0.08$ when they 
keep only the $K \Lambda$ intermediate state.

In the extreme scenario that the whole Karliner 
``discrepancy'' were attributed to $F_2^s(0)$
one would predict $F_2^s \simeq +0.66$, which 
is considerably bigger 
and has 
the opposite sign to most theoretical calculations of $F_2^s(0)$.
Of course, the non-strange part of the kaon loop
$N \rightarrow K \Lambda$ 
also renormalises $\kappa_N$.
To the best of our knowledge, this kaon loop contribution has not 
yet been
calculated.
Further contributions will come from 
$\gamma \Lambda \rightarrow \Lambda^*$ 
excitations 
inside the kaon loop.
A large positive value of $F_2^s(0)$ ($\simeq +{1 \over 2}$) 
would also follow if the SU(3) flavour structure of 
the anomalous magnetic moment $\kappa_N$ reflects the up quark 
dominance of 
many low-energy nucleon observables in QCD, 
viz. if
$|F_2^u(0)| > |F_2^d(0)|$ \cite{mcke}.
(The physical values of $\kappa_N$, Eq.(21), 
 together with $F_2^s(0) = +0.5$, would imply $F_2^u(0) = +2.17$ 
 and $F_2^d(0) = -1.53$.)

Given the important role that the nucleon resonances play in
Karliner's evaluation of the sum-rule, it 
seems reasonable to consider the electromagnetic excitations 
of the $\Lambda$ in the processes
($N \rightarrow K \Lambda$, 
$\gamma \Lambda \rightarrow \Lambda^*$).
The interference between this process,
which is associated with the isoscalar anomalous magnetic moment,
and
the process
($\gamma N \rightarrow N^*$, $N^* \rightarrow K \Lambda^*$), 
which is, in part, associated with the isovector anomalous 
magnetic moment, 
is a
candidate to fill the Karliner ``discrepancy''.
In Table 3 we list the threshold energies $\nu_{\rm LAB}^{th}$ for
the photoproduction of the $\Lambda$ and 
$\Sigma$ hyperons, together with their first two excited resonances 
\cite{pdg}.
\begin{table}
\begin{center}
\caption{Thresholds for hyperon (resonance) production in a DHG
  experiment}
\begin{tabular} {cccc}
\\
\hline\hline
\\
Hyperon              &  $J^{P}$          &  Resonance     &  $\nu_{\rm LAB}^{th}$  \\
\hline
\\
$\Lambda(1115)$      &  ${1 \over 2}^+$  &                &  0.91 GeV    \\
$\Lambda^{*}(1405)$  &  ${1 \over 2}^-$  &  $S_{01}$      &  1.45 GeV    \\
$\Lambda^{*}(1520)$  &  ${3 \over 2}^-$  &  $D_{03}$      &  1.69 GeV    \\
\\
$\Sigma^+(1189)$     &  ${1 \over 2}^+$  &                &  1.05 GeV    \\
$\Sigma^0(1193)$     &  ${1 \over 2}^+$  &                &  1.05 GeV    \\
$\Sigma^{*}(1385)$   &  ${3 \over 2}^+$  &  $P_{13}$      &  1.42 GeV    \\
$\Sigma^{*}(1680)$   &  ${1 \over 2}^+$  &  $P_{11}$      &  2.06 GeV    \\
\\
\hline\hline
\end{tabular}
\end{center}
\end{table}
One priority for future polarised photoproduction experiments
should be to reconstruct these 
$\gamma N \rightarrow K \Lambda^*$ 
and $\gamma N \rightarrow K \Sigma^*$
processes from events involving 
one or two kaons in the final state.

Whilst this article was being prepared, Hammer et al.\cite{hamm}
independently suggested investigating exclusive channels involving 
strange mesons ($K$, $\eta$ and $\phi$)
as a possible source of the Karliner ``discrepancy''.
Using positivity arguments, they argue that unpolarised single
$K$, $\eta$ and $\phi$ photoproduction data puts an upper bound
$|\kappa_{\rm HDM}| \leq 0.2 \kappa_p$ on the contribution 
$\kappa_{\rm HDM}$
of these processes to the value of the anomalous magnetic moment
that is extracted from the Drell-Hearn-Gerasimov sum-rule.

\section{$Q^2$ dependence of $(\sigma_A - \sigma_P)$}

The $Q^2$ and $\nu$ dependence of 
$(\sigma_A - \sigma_P)$ is contained in two spin dependent nucleon 
form factors 
$G_1(\nu, Q^2)$ and $G_2(\nu, Q^2)$, viz.
\begin{equation}
\sigma_A - \sigma_P =
{16 m \pi^2 \alpha \over 2 m \nu - Q^2} 
\Biggl( 
m \nu G_1 (\nu, Q^2) - Q^2 G_2 (\nu, Q^2) 
\Biggr).
\end{equation}
Following Anselmino, Ioffe and Leader \cite{ansel}, the $Q^2$
dependent quantity
\begin{equation}
I (Q^2) = m^3 \int_{ {Q^2 \over 2m} }^{\infty} {d \nu \over \nu} 
G_1 (\nu, Q^2)
\end{equation}
can be used to interpolate between polarised photoproduction and
deep inelastic scattering. The Drell-Hearn-Gerasimov sum-rule Eq.(1) 
is the statement:
\begin{equation}
I(0) = - {1 \over 4} \kappa^2_N.
\end{equation}
In high $Q^2$ deep inelastic scattering 
\begin{equation}
m^2 \nu G_1 (\nu, Q^2) \rightarrow g_1 (x, Q^2), \ \ \ 
m \nu^2 G_2 (\nu, Q^2) \rightarrow g_2 (x, Q^2)
\end{equation}
where $x= {Q^2 \over 2m \nu}$ is the Bjorken variable.
In the deep inelastic limit
\begin{equation}
I (Q^2) =
{2 m^2 \over Q^2}
\int_0^1 dx g_1 (x, Q^2).
\end{equation}
As $Q^2$ tends to $\infty$, the first moment of $g_1$
is given by 
\cite{mink,kod,larin,altr}
\begin{equation}
\int^1_0 dx g_1(x,Q^2) 
= {1 \over 2} \sum_{\rm q} {\rm e_q^2} \Delta q_{\rm inv}
\Bigl\{1 + \sum_{\ell\geq 1} c_\ell\,\bar{g}^{2\ell}(Q)\Bigr\}
+ {\cal O}\Bigl({1 \over Q^2}\Bigr)
\end{equation}
Here 
$\Delta q_{\rm inv}$ is the q flavoured, 
gauge and scale
invariant axial charge \cite{bcft}.
The coefficient $c_\ell$ is calculable in 
$\ell$-loop perturbation theory \cite{larin}.
The current experimental value of the singlet axial
charge
\begin{equation}
g_A^0|_{\rm inv} = \Delta u_{\rm inv} + \Delta d_{\rm inv} +
\Delta s_{\rm inv}
\end{equation}
is \cite{expt, badelek}
\begin{equation}
g_A^0|_{\rm inv} = 0.28 \pm 0.07.
\end{equation}
This number is four standard deviations below the value 0.58
that it would have taken if 
$\Delta s_{\rm inv}$ were zero \cite{ej}.

The large violation of OZI in $g_A^0|_{\rm inv}$ 
is commonly associated
with the $U_A(1)$ axial anomaly in QCD [18-23].
The axial anomaly does not contribute to the anomalous magnetic
moment
for the reasons outlined in Section 4.
This source of OZI violation gives a vanishing contribution
to the Drell-Hearn-Gerasimov integral 
$I(0)$ at photoproduction.
Strange resonance excitations
$\gamma N \rightarrow K \Lambda^*$, 
and
$\gamma N \rightarrow K \Sigma^*$
have the potential to make an important contribution to $I(0)$
at photoproduction.
These resonance excitations contribute to polarised deep inelastic
scattering
only at non-leading twist.
They are not present in $g_A^0|_{\rm inv}$.

Several models [57,79-86] have been proposed for the non-leading
twist $Q^2$ variation of $I(Q^2)$, 
which changes sign between polarised photoproduction, 
Eq.(28), and deep inelastic scattering, Eqs.(31-33).
Bag model \cite{ji} and QCD sum-rule \cite{bbk,stein} 
calculations are available for the twist four 
${\cal O}({1 \over Q^2})$ corrections to $g_1$ 
in Eq.(31).
These calculations do not yet include the physics of the axial 
anomaly, which is 
potentially very important in the flavour singlet channel \cite{ioffe}.
Ioffe and collaborators \cite{ansel, iof92} have proposed a
simple phenomenological model for the $Q^2$ dependence of 
the inelastic part of
$(\sigma_A - \sigma_P)$.
They argue that the contribution from resonance production,
denoted $I^{\rm res}(Q^2)$,
has a strong $Q^2$ dependence for small $Q^2$ and then drops
rapidly with $Q^2$.
The non-resonant part of $I(Q^2)$ is then parametrised by a 
smooth function which they took as the sum of a monopole and 
a dipole term, viz.
\begin{equation}
I(Q^2) = 
I^{\rm res} (Q^2) + 2 m^2 \Gamma^{\rm as}
\Biggl( {1 \over Q^2 + \mu^2} 
- {C \mu^2 \over (Q^2 + \mu^2)^2} \Biggr).
\end{equation}
Here 
\begin{equation}
\Gamma^{\rm as} = \int_0^1 dx g_1 (x, \infty)
\end{equation}
and 
\begin{equation}
C = 1 + {1 \over 2} 
\ {\mu^2 \over m^2} \ {1 \over \Gamma^{\rm as}} \
\Bigl( {1 \over 4} \kappa^2 + I^{\rm res}(0) \Bigr).
\end{equation}
Using vector meson dominance arguments, the mass parameter 
$\mu$ was identified with rho meson mass, $\mu^2 \simeq m_{\rho}^2$.
Karliner's analysis \cite{ikar} suggests that 
$I^{\rm res}_p(0) = -1.01$.
(There is no elastic contribution to (DHG).)
In this approach,
the ``discrepancy'' in the Drell-Hearn-Gerasimov sum-rule
$(I - I^{\rm res})_p(0) = +0.21$
is associated with
the non-resonant vector meson dominance term.

The E143 Collaboration at SLAC have recently announced the first
low-$Q^2$ measurements of $I_p(Q^2)$ and $I_d(Q^2)$ at 
$Q^2 = 0.5$GeV$^2$ and  $Q^2 = 1.2$GeV$^2$ \cite{e143}.
Both the proton and deuteron data points are consistent 
with the predictions 
of \cite{iof92, soff}.
The $Q^2 = 1.2$GeV$^2$ data is consistent with the QCD 
sum-rule calculations \cite{bbk, stein} of the twist 
four contribution to Eq.(31) with coefficients 
$c_\ell$ which are evaluated 
to ${\cal O}(\alpha_s^3)$ \cite{larin, alta}.
(Theoretical uncertainties due to higher order coefficients 
 may be difficult to estimate \cite{mart}.)

\section {Conclusions and Opportunities}

Photoproduction spin sum-rules offer a new window on the
spin structure of the nucleon that complements the
information that we can learn from polarised deep inelastic
scattering experiments.
The first direct measurements of the spin photoproduction
cross-sections
$\sigma_A$ and $\sigma_P$ will soon be available from the
ELSA, GRAAL, LEGS and MAMI facilities up to $\sqrt{s} \leq 2.5$GeV.
These experiments will make a precise check of $\gamma N \rightarrow
N^{*}$ resonance contributions to the Drell-Hearn-Gerasimov and Spin 
Polarisability sum-rules.
Further important information on the photoproduction spin structure 
of the nucleon will come from measuring:
\begin{itemize}
\item
Non-resonant contributions associated, in part, with vector 
meson dominance (Section 6) and
\item
Strange resonances 
($\gamma N \rightarrow K \Lambda^{*}$) 
and
($\gamma N \rightarrow K \Sigma^{*}$)
as possible sources
of the Karliner ``discrepancy'' (Sections 4 and 5);
\item
Tests of spin dependent Regge theory to distinguish 
the Lorentz structure of pomeron-like exchanges Eqs.(12-14).
This requires varying $\sqrt{s} \geq 10$GeV
to ensure that we are in the region
where the pomeron dominates over other Regge contributions
\cite{pvl1} (Section 2).
Regge behaviour provides a good description of 
$(\sigma_A + \sigma_P)$  starting at $\sqrt{s} \simeq 2.5$GeV.
\end{itemize}
When combined with measurements of the $Q^2$
dependence of
$(\sigma_A - \sigma_P)$ from SLAC and TJNAF, we will soon have a 
much more complete picture of the nucleon's 
internal
spin structure.

\pagebreak
\vspace{1.0cm}
{\large \bf Acknowledgements: \\}
\vspace{3ex}

It is a pleasure to thank S.J. Brodsky, R.J. Crewther, N. d'Hose, 
B. Metsch, W. Melnitchouk, H. Petry, B. Schoch, D. Sch\"utte and 
A.W. Thomas for helpful discussions on various aspects of DHG 
physics. This work was supported by a Research Fellowship of the 
Alexander von Humboldt Foundation.



\begin{thebibliography}{99}
%
\bibitem{dhg}
S.D. Drell and A.C. Hearn, Phys. Rev. Lett. 162 (1966) 1520; \\
S.B. Gerasimov, Yad. Fiz. 2 (1965) 839.
%
\bibitem{brod69}
S.J. Brodsky and J.R. Primack, Ann. Phys. 52 (1969) 315.
%
\bibitem{alt}
G. Altarelli, N. Cabibbo and L. Maiani, Phys. Lett. B 40
    (1972) 415.
%
\bibitem{brodsc}
S.J. Brodsky and I. Schmidt, Phys. Lett. B 351 (1995) 344.
%
\bibitem{drell}
S.J. Brodsky and S.D. Drell, Phys. Rev. D 22 (1980) 2236.
%
\bibitem{rizzo}
T.G. Rizzo, Phys. Rev. D54 (1996) 3057.
%
\bibitem{bj} J.D. Bjorken, Phys. Rev. 148 (1966) 1467; Phys. Rev.
D 1 (1970) 1376.
%
\bibitem{ej} J. Ellis and R.L. Jaffe, Phys. Rev. D 9 (1974) 1444;
(E) D 10 (1974) 1669.
%
\bibitem{mink} P. Minkowski, {\em in\/} Proc. Workshop on {\em
Effective Field Theories of the Standard Model\/}, Dobog\'{o}k\~{o},
Hungary 1991, ed. U.-G. Meissner (World Scientific, Singapore, 1992).
%
\bibitem{kod} J. Kodaira, Nucl. Phys. B 165 (1980) 129.
%
\bibitem{larin} S.A. Larin, Phys. Lett. B 334 (1994) 192.
%
\bibitem{altr}
G. Altarelli and G. Ridolfi, Nucl. Phys. B (Proc. Suppl.) 
39B,C (1995) 106.
%
\bibitem{bcft}
S.D. Bass, R.J. Crewther, F.M. Steffens and A.W. Thomas, 
hep-ph/9701213.
%
\bibitem{exptbj}
SMC Collaboration,
D. Adams et al., Phys. Lett. B 357 (1995) 248; \\
E143 Collaboration, 
K. Abe et al., Phys. Rev. Lett. 74 (1995) 346.
\bibitem{abfr}
G. Altarelli, R.D. Ball, S. Forte and G. Ridolfi, hep-ph/9701289.
%
\bibitem{expt} EMC Collaboration, J Ashman et al., Phys. Lett. B
  206(1988) 364; 
Nucl. Phys. B 328 (1989) 1; \\
SMC Collaboration, D. Adams et al., Phys. Lett. B 329 (1994) 399;
(E) B 339 (1994) 332; \\
E143 Collaboration, K. Abe et al., Phys. Rev. Lett. 75 (1995) 25. 
%
\bibitem{badelek} B. Bade{\l}ek, CRAD96 conference, Cracow, August
1996 (hep-ph/9612274).
%
\bibitem{efremov} A.V. Efremov and O.V. Teryaev, JINR Report
E2--88--287 (1988), and in Proceedings of the International Hadron
Symposium, Bechyn\v{e} 1988, eds.\ J. Fischer et al.\ 
(Czechoslovakian Academy of Science, Prague, 1989) p. 302;\\
G. Altarelli and G.G. Ross, Phys. Lett. B 212 (1988) 391;\\
R.D. Carlitz, J.C. Collins, and A.H. Mueller, Phys. Lett. B 214
(1988) 229.
%
\bibitem{jaffea} R.L. Jaffe and A. Manohar, Nucl. Phys. B 337 (1990)
509;\\
R.L. Jaffe, Phys. Lett. B 365 (1996) 359.
%
\bibitem{venez}
G. Veneziano, Mod. Phys. Lett. A 4 (1989) 1605; \\
G.M. Shore and G. Veneziano, Nucl. Phys. B 381 (1992) 23.
%
\bibitem{fort}
S. Forte, Nucl. Phys. B 331 (1990) 1.
%
\bibitem{frit}
H. Fritzsch, Phys. Lett. B 256 (1991) 75.
%
\bibitem{bass}
S.D. Bass, Z Physik C 60 (1993) 343.
%
\bibitem{bek}
S.J. Brodsky, J. Ellis and M. Karliner, Phys. Lett. B 206 (1988) 309.
%
\bibitem{cheng}
H.-Y. Cheng, Int. J. Mod. Phys. A 11 (1996) 5109; \\
J. Ellis, hep-ph/9611208.
%
\bibitem{ikar}
I. Karliner, Phys. Rev. D 7 (1973) 2717. 
%
\bibitem{ffox}
G.C. Fox and D.Z. Freedman, Phys. Rev. 182 (1969) 1628.
%
\bibitem{drec}
D. Drechsel, Prog. Part. Nucl. Phys. 34 (1995) 181.
%
\bibitem{low}
F. Low, Phys. Rev. 96 (1954) 1428; \\
M. Gell-Mann and M.L. Goldberger, Phys. Rev. 96 (1954) 1433.
%
\bibitem{coll}
P.D.B. Collins, {\it Regge Theory and High Energy Physics}, 
Cambridge UP (1978).
%
\bibitem{peie}
R.F. Peierls and T.L. Trueman, Phys. Rev. 134 (1964) 1365; \\
A.H. Mueller and T.L. Trueman, Phys. Rev. 160 (1967) 1296, 1306.
%
\bibitem{heim}
R.L. Heimann, Nucl. Phys. B 64 (1973) 429.
%
\bibitem{ek}
J. Ellis and M. Karliner, Phys. Lett. B 213 (1988) 73.
%
\bibitem{clos1}
F.E. Close and R.G. Roberts, Phys. Lett. B 336 (1994) 257.
%
\bibitem{pvl1}
P.V. Landshoff, 
Proc. Zuoz Summer School, PSI Proceedings 94-01 (1994) 135,
hep-ph/9410250.
%
\bibitem{pvl2}
P.V. Landshoff and O. Nachtmann, Z Physik C 35 (1987) 405. 
%
\bibitem{sbpl}
S.D. Bass and P.V. Landshoff, Phys. Lett. B 336 (1994) 537.
%
\bibitem{kuti}
L. Galfi, J. Kuti and A. Patkos, Phys. Lett. B 31 (1970) 465; \\
F.E. Close and R.G. Roberts, Phys. Rev. Lett. 60 (1988) 1471.
%
\bibitem{dglap}
R.D. Ball, S. Forte and G. Ridolfi, 
Nucl. Phys. B444 (1995) 287; (E) B449 (1995) 680.
%
\bibitem{bfkl}
J. Bartels, B.I. Ermolaev and M.G. Ryskin, Z Phys C70 (1996) 273; 
C72 (1996) 627.
%
\bibitem{Muta}
T. Muta, {Foundations of Quantum Chromodynamics} (World Scientific,
1987); \\
G. Altarelli, Phys. Rep. 81 (1982) 1.
%
\bibitem{dash}
S.L. Adler and R.F. Dashen, {\it Current Algebras and Applications
to Particle Physics} (W.A. Benjamin, 1968).
%
\bibitem{brod72}
S.J. Brodsky, F.E. Close and J.F. Gunion, Phys. Rev. D 5 (1972) 1384.
%
\bibitem{bgj} 
D.J. Broadhurst, J.F. Gunion and R.L. Jaffe, Ann. Phys. 81 (1973) 88.
%
\bibitem{pfeil}
R. Pantf\"order, H. Rollnik and W. Pfeil, hep-ph/9703368.
%
\bibitem{gold}
H.D. Abarbanel and M.L. Goldberger, Phys. Rev. 165 (1968) 1594.
%
\bibitem{adler} S.L. Adler, Phys. Rev. 177 (1969) 2426.
%
\bibitem{bell} J.S. Bell and R. Jackiw, Nuovo Cimento 60 A (1969)
47.
%
\bibitem{moi} 
S. D. Bass, Z Physik A 355 (1996) 77.
%
\bibitem{gilm}
M. Damashek and F.J. Gilman, Phys. Rev. D 1 (1970) 1319.
%
\bibitem{cloud}
A. W. Thomas, Adv. Nucl. Phys. 13 (1984) 1; \\
S. Th\'eberge, G.A. Miller and A.W. Thomas, Can. J. Phys. 60 (1982) 59.
%
\bibitem{schl}
S.J. Brodsky and F. Schlumpf, Phys. Lett. B 329 (1994) 111.
%
\bibitem{drec92}
D. Drechsel and L. Tiator, J. Phys. G 18 (1992) 44.
%
\bibitem{workm}
R.L. Workman and R.A. Arndt, Phys. Rev. D 45 (1992) 1789.
%
\bibitem{spinp}
A.M. Sandorfi, C.S. Whisnant and M. Khandaker, Phys. Rev. D 50 (1994) R6681.
%
\bibitem{weise}
N. Kaiser, T. Waas and W. Weise, Nucl. Phys. A 612 (1997) 297.
%
\bibitem{chpt}
V. Bernard, N. Kaiser, J. Kambor and U.G. Meissner, Nucl. Phys. B 388
(1992) 315; \\
V. Bernard, N. Kaiser and U.G. Meissner, Phys. Rev. D 48 (1993) 3062.
%
\bibitem{jaffe}
R.L. Jaffe, Phys. Lett. B 229 (1989) 275; \\
H.W. Hammer, U.G. Meissner and D. Drechsel, Phys. Lett. B 367 (1996)
323.
%
\bibitem{meis97}
U-G. Meissner, V. Mull, J. Speth and J.W. van Orden, hep-ph/9701296.
%
\bibitem{fork}
T.D. Cohen, H. Forkel and M. Nielsen, Phys. Lett. B 316 (1993) 1; \\
H. Forkel, M. Nielsen, X. Jin and T.D. Cohen, Phys. Rev. C 50 (1994)
3108; \\
H. Forkel, hep-ph/9607452.
%
\bibitem{henl}
W. Koepf, E.M. Henley and S.J. Pollock, Phys. Lett. B 288 (1992) 11.
%
\bibitem{muso}
M.J. Musolf and M. Burkardt, Z Phys. C 61 (1994) 433.
%
\bibitem{wall}
W. Melnitchouk and M. Malheiro, Phys. Rev. C 55 (1997) 431.
%
\bibitem{isgur}
P. Geiger and N. Isgur, Phys. Rev. D 55 (1997) 299.
%
\bibitem{mus96}
M.J. Musolf and H. Ito, nucl-th/9607021.
%
\bibitem{goeke}
H-C. Kim, T. Watabe and K. Goeke, hep-ph/9606440.
%
\bibitem{parka}
N.W. Park, J. Schechter and H. Weigel, Phys. Rev. D 43 (1991) 869.
%
\bibitem{parkb}
S.T. Hong and B. Park, Nucl. Phys. A 561 (1993) 525.
%
\bibitem{lein}
D. Leinweber, Phys. Rev. D 53 (1996) 5115.
%
\bibitem{hohl1}
G. H\"ohler et al., Nucl. Phys. B 224 (1976) 505.
%
\bibitem{hohl2}
H. Genz and G. H\"ohler, Phys. Lett. B 61 (1976) 389.
%
\bibitem{holi}
G. Jansen, K. Holinde and J. Speth, Phys. Rev. Lett. 73 (1994) 1332.
%
\bibitem{brod80}
S.J. Brodsky and G.R. Farrar, Phys. Rev. D 11 (1977) 1309; \\
G. P. Lepage and S. J. Brodsky, Phys. Rev. D 22 (1980) 2157.
%
\bibitem{sign}
O. Dumbrajs et al., Nucl. Phys. B216 (1983) 277; \\
A.I. Signal and A.W. Thomas, Phys. Lett. B 191 (1987) 205.
%
\bibitem{shimoda}
A. W. Thomas and W. Melnitchouk, in {\it New  Frontiers in Nuclear
Physics},
Eds. S. Homma, Y. Akaishi and M. Wada (World Scientific, Singapore, 1993),
pp. 41-106.
%
\bibitem{mcke}
R.D. McKeown, hep-ph/9607340.
%
\bibitem{pdg}
Particle Data Group, M. Aguilar-Benitez et al., Phys. Rev. D 50 (1994)
1173.
%
\bibitem{hamm}
H.W. Hammer, D. Drechsel and T. Mart, nucl-th/9701008.
%
\bibitem{ansel}
M. Anselmino, B. L. Ioffe and E. Leader, Yad. Fiz. 49 (1989) 214.
%
\bibitem{ji}
X. Ji, Phys. Lett. B 309 (1993) 187; \\
X. Ji and P. Unrau, Phys. Lett. B 333 (1994) 228 \\
X. Ji and W Melnitchouk, hep-ph/9703363. 
%
\bibitem{bbk}
I.I. Balitsky, V.M. Braun and A.V. Kolesnichenko, Phys. Lett. B 242
(1990) 245; (E) B 318 (1993) 648.
%
\bibitem{stein}
E. Stein et al., Phys. Lett. B 343 (1995) 369; B 353 (1995) 107.
%
\bibitem{iof92}
V.D. Burkert and B.L. Ioffe, Phys. Lett. B 296 (1992) 223; \\
V.D. Burkert and B.L. Ioffe, JETP 78 (1994) 619; \\
B. L. Ioffe, hep-ph/9704295.
%
\bibitem{soff}
J. Soffer and O. Teryaev, Phys. Rev. Lett. 70 (1993) 3373;
Phys. Rev. D 51 (1995) 25; hep-ph 9703447.
%
\bibitem{burk}
V. Burkert and Zh. Li, Phys. Rev. D 47 (1993) 46.
%
\bibitem{li}
Z-P. Li and Zh. Li, Phys. Rev. D 50 (1994) 3119.
%
\bibitem{ioffe}
B.L. Ioffe, Erice Lecture, hep-ph/9511401.
%
\bibitem{e143} 
E143 Collaboration, K. Abe et al., Phys. Rev. Lett. 78 (1997) 815.
%
\bibitem{alta}
G. Altarelli, P. Nason and G. Ridolfi, Phys. Lett. B 320 (1994) 152;
(E) B 325 (1994) 538.
%
\bibitem{mart}
G. Martinelli and C.T. Sachrajda, Nucl. Phys. B 478 (1996) 660.
%
\end{thebibliography}
\end{document}